\documentclass[ams, prb, twocolumn, amssymb,floatfix]{revtex4-1}
\usepackage{amsmath}
\usepackage{tabularx}
\usepackage{bm}
\usepackage{euscript}
\usepackage{graphicx}
\usepackage{color}
\usepackage{amsfonts}
\usepackage{exscale}
\usepackage{amsbsy}
\usepackage{subfigure}
\usepackage{textcomp}
\usepackage{comment}
\usepackage{slashed}
\usepackage{hyperref}

\newcommand{\beq}{\begin{equation}}
\newcommand{\eeq}{\end{equation}}
\newcommand{\be}{\begin{eqnarray}}
\newcommand{\ee}{\end{eqnarray}}

\begin{document}

\title{Emergent reflection symmetry from non-relativistic composite fermions}
\author{Prashant Kumar$^{1}$, Michael Mulligan$^{2}$, S. Raghu$^{1,3}$}
\affiliation{$^{1}$Stanford Institute for Theoretical Physics, Stanford University, Stanford, California 94305, USA}
\affiliation{$^{2}$Department of Physics and Astronomy, University of California, Riverside, Riverside, CA 92511, USA}
\affiliation{$^3$SLAC National Accelerator Laboratory, 2575 Sand Hill Road, Menlo Park, CA 94025, USA}
\date{\today}

\begin{abstract}
A recent experimental study [Pan et al., arXiv: 1902.10262] has shown that fractional quantum Hall effect gaps are essentially consistent with particle-hole symmetry in the lowest Landau level.  Motivated by this result, we consider a clean two dimensional electron system (2DES) from the viewpoint of composite fermion mean-field theory.  In this short note, we show that while the experiment is manifestly consistent with a Dirac composite fermion theory proposed recently by Son, it can equally well be explained within the framework of non-relativistic composite fermions, first put forward by Halperin, Lee, and Read.  
\end{abstract}

\maketitle

\section{Introduction}

In the limit of an infinitely strong perpendicular magnetic field, the dynamics of a two dimensional electron system (2DES) is governed entirely by the lowest Landau level (LLL). \cite{prange1987quantum, dassarmaqhereview} In the absence of disorder, it is well known that electrons in the LLL interacting with Coulomb forces explicitly satisfy particle-hole symmetry. \cite{PhysRevLett.50.1219, girvin1984} For example, the properties of fractional quantum Hall states at filling fraction $\nu$ are identical to those at filling fraction $1- \nu$, since both are particle-hole conjugates of one another.  To the extent that particle-hole symmetry breaking effects, such as quenched disorder and  Landau level mixing are weak, particle-hole symmetric response can be observed directly in an experiment.

In a recent study,\cite{Pan2019} magnetotransport measurements of a high mobility 2DES  have directly verified the expectation of particle-hole symmetry.  The 2DES was studied at a fixed external magnetic field, while the chemical potential was tuned (through electrostatic gating), enabling a comparative study of fractional quantum Hall states at $\nu$ and $1- \nu$ at the same magnetic field.  The authors reported that gaps at $\nu= 1/3$ and $\nu=2/3$, extracted from the temperature dependence of the longitudinal resistance,  were nearly identical.  Similar observations occur between particle-hole conjugate states at $\nu$ and $1-\nu$, for  $\nu= 2/5, 3/7, 4/9$.  In addition, the experiments in Ref.~\onlinecite{Pan2019} also display a reflection symmetry in the conductivity tensor about $\nu=1/2$, which implies particle-hole symmetric electromagnetic response within the LLL.  

While the existence of particle-hole symmetry is well-understood from microscopic physics in the lowest Landau level, its realization in low-energy effective descriptions is challenging.  A prominent such theory involves emergent fermionic particles known as composite fermions, which roughly correspond to electrons bound to two flux quanta. \cite{Jainbook, Fradkinbook} Such theories have successfully explained the phenomenon of a gapless Fermi liquid like state at $\nu=1/2$.  At filling fraction $\nu=1/2$, composite fermions encounter, on average,  zero net magnetic field and fill up a Fermi sea.  Neighboring fractional quantum Hall states away from $\nu=1/2$ correspond to integer quantum Hall states of composite fermions.  This way, much of the phenomenology of the quantum Hall effect can be captured qualitatively by composite fermion descriptions.\cite{Willett97}  One remaining challenge, which has been the focus of a great deal of recent work,\cite{Son2015, BMF2015, Geraedtsetal2015, BalramRifmmodeCsabaJain2015, PotterSerbynVishwanath2015, PhysRevB.94.245107, mulliganraghufisher2016, LevinSon2016, 2017PhRvX...7c1029W, 2018arXiv180307767K, PhysRevB.98.115105} amounts to determining how the constraint of particle-hole symmetry can be incorporated in composite fermion based theories.\cite{kivelson1997} 

In this note, we consider the observation of particle-hole symmetric fractional quantum Hall states within various composite fermion descriptions.  We consider a non-relativistic theory\cite{halperinleeread} of composite fermions first studied by Halperin, Lee and Read (HLR).  While the HLR Lagrangian appears to violate particle-hole symmetry, we show how nevertheless the theory is perfectly capable of exhibiting particle-hole symmetric electromagnetic response.  

\section{Statement of the problem}
In this section, we phrase the constraint of the experiments in Ref.~\onlinecite{Pan2019}.  Electrons in the lowest Landau level are described by a Lagrangian density of the form
\begin{equation}
\mathcal L = \bar \psi \left[\hat K_A + \mu \right] \psi + \ldots,
\end{equation}
where $\hat K_A = i D_A^t  + \frac{1}{2m} \vec D_A^2$, $D^\mu_{A} = \partial_{\mu} - i A_{\mu}, \mu \in \{t,x,y\}$, and the $\ldots$ are crucial interactions that lift the extensive Landau level degeneracy, giving rise to the fractional quantum Hall effect.  At filling fraction $\nu$, the chemical potential $\mu$ is adjusted such that 
\begin{equation}
\langle \bar \psi \psi \rangle = \nu \frac{B}{2 \pi}, \ \ B = \partial_x A_y - \partial_y A_x.
\end{equation}
The composite fermion framework gets around the challenge of contending with the degeneracy, by making an exact mapping to a system of composite fermions coupled to both the external gauge field, and a dynamical $U(1)$ gauge field with a Chern-Simons (CS) term:
\begin{equation}
\label{HLR}
    \mathcal{L} = \bar f \left[\hat K_{a + A} + \mu \right] f + \frac{1}{8 \pi} ada + \cdots
\end{equation}
where $ada \equiv \epsilon_{\mu \nu \lambda} a_{\mu} \partial_{\nu} a_{\lambda}$, and the $\cdots$ now also include a Maxwell term for $a$ and less relevant couplings.  The above theory is as difficult to analyze as the electron theory; its advantage, however, is that it motivates a simple mean-field approximation that enables us to capture the essence of the relevant physics.  We will restrict our analysis exclusively to a mean-field treatment of the composite fermion system.  Moreover, since our interest here is only in the gapped fractional quantum Halls states, such mean-field approximations provide leading order terms of a series with finite radius of convergence.  

Having disposed of such caveats, let us focus on the issue of particle-hole symmetry.  From the equation of motion of $a_t$, namely,
\begin{equation}
\langle \bar f f \rangle + \frac{b}{4 \pi} = 0, \ \ b = \partial_x a_y - \partial_y a_x 
\end{equation}
and since the composite fermions and electrons are at the same density, 
\begin{equation}
\langle \bar \psi \psi \rangle = \langle \bar f f \rangle  \Rightarrow   2 \nu B + b = 0
\end{equation}
The significance of the above expression is that since the composite fermions couple to $a+A$, the net magnetic field experienced by composite fermions in mean-field theory is $b + B$.  At $\nu = 1/2$, this field vanishes, and the mean-field ground state is a filled Fermi sea of composite fermions.  

Let us define a composite fermion filling fraction $\nu_{cf}$ as the ratio of the density of composite fermions and the flux density of the total field $b + B$.  From the expressions above, it follows,
\begin{equation}
\nu_{cf} = 2 \pi \frac{\langle \bar f f \rangle}{b+B} = \frac{\nu}{1-2 \nu}.
\end{equation}
Thus, while $\nu=1/3$ corresponds to $\nu_{cf} = 1$, the particle-hole conjugate state $\nu=2/3$ corresponds to $\nu_{cf} = -2$.  Since the $\nu_{cf} = 1$ and $\nu_{cf} = -2$ states are not related by any obvious underlying symmetry, it seems impossible for these two states to make equivalent predictions for the electronic systems, unless there were significant fine-tuning at play.  It would seem, therefore, that the HLR theory cannot make contact with the experiment described above.  Such a hasty declaration, however, is false, as we describe next.  

\section{Resolution of the problem\label{resolution_section}}
A careful analysis of the equations of motion of Eq. \eqref{HLR} leads to a natural resolution of the puzzle posed in the previous section.  Consider the following thought experiment.  Let us start with the chemical potential $\mu$ tuned such that $\nu = 1/2$, and the density of electrons and composite fermions is $n_{1/2} = B/4\pi$.  Next, to access nearby fractional quantum Hall states, we vary $\mu$.  One must be careful to note, however, that as we vary $\mu$, we change the density of composite fermions, which in turn affects the density of flux $b+B$ encountered by composite fermions.  To be explicit, let $\mu \rightarrow \mu + V$, where $V$ is the change in the gate voltage.  Within linear response, the corresponding change in the density of composite fermions is related to the change in chemical potential via the compressibility $\chi = m/2 \pi$ of composite fermions,\cite{KimFurusakiWenLee1994}
\begin{equation}
\langle \bar f f \rangle = \frac{B}{4 \pi} + \frac{m}{2 \pi} V = -\frac{b}{4 \pi},
\end{equation}
where we used the equation of motion of $a_t$ in the second equality above.  Therefore, we may relate the shift in the chemical potential to an effective magnetic field
\begin{equation}
\label{constraint}
V = -\frac{B+b}{2m}.
\end{equation}
We can then incorporate this constraint in Eq.~\eqref{constraint} and write the HLR Lagrangian in the presence of $V$ as
\begin{equation}
\label{constrained}
\mathcal{L} = \bar f \left[ \hat K_{a+A} + \mu  - \frac{b+B}{2m} \right] f + \frac{1}{8 \pi} ada + \ldots
\end{equation}
Observing again that the composite fermions only couple to the sum $a+A$, we may shift the dynamical field $a \mapsto a - A$, and rewrite the above Lagrangian as
\begin{align}
\label{geq2}
\mathcal{L} = \bar f \left[ \hat K_{a} + \mu  - \frac{b}{2m} \right] f + \frac{1}{8 \pi} \left(a-A \right) d \left(a-A \right) + \ldots
\end{align}
In a mean-field approximation, the first term above describes a non-interacting particle in a magnetic field $b = \partial_x a_y - \partial_y a_x$ with an additional Zeeman-like coupling $-\frac{b}{2m} \bar f f$.  The corresponding Landau level spectrum is
\begin{equation}
E_n = \frac{\vert b \vert}{m} \left( n + \frac{1+\zeta}{2}  \right), \ \ \ n = 0, 1, 2, \ldots
\end{equation}
and where $\zeta = {\rm sgn}(b)$ depends on the sign of the composite fermion magnetic field, which, in the original electron coordinates determines whether the electron filling fraction is being increased or decreased from $\nu = 1/2$.  The key observation to make above is that with a fixed $\mu$, the number of filled composite fermion Landau levels depends on $\zeta = {\rm sgn}(b)$: if $p$ Landau levels are filled for $\zeta = 1$, then $p+1$ Landau levels are filled for the same $\mu$ when $\zeta = -1$.  This mismatch of one Landau level precisely corrects the discrepancy discussed in the previous section.  

To see how this comes about, imagine that the chemical potential is tuned to a fractional quantum Hall plateau, with some integer $p$ of composite fermion Landau levels being filled for $\zeta = 1$.  Then, upon integrating out the composite fermions, we find  
\begin{equation}
\bar f \left[ \hat K_a + \mu  - \frac{b}{2m} \right] f = \frac{p + \frac{1-\zeta }{2}}{4 \pi} a da +  \ldots
\end{equation}
where $\ldots$ now refer to subleading corrections that are suppressed by the fractional quantum Hall energy gaps.  
Gathering both terms in Eq.~\eqref{geq2}, we find
\begin{equation}
\mathcal L = \zeta \frac{p + \frac{1-\zeta }{2}}{4 \pi} a da + \frac{1}{8 \pi} \left(a-A \right) d \left(a-A \right) + \ldots
\end{equation}
To obtain the electromagnetic response, we integrate out $a$ and obtain
\begin{eqnarray}
\mathcal L^{\rm eff}_{b>0} &=& \frac{1}{4 \pi} \frac{p}{2 p + 1}AdA,  \ \ \ \ \nu = \frac{p}{2p+1} \nonumber \\
\mathcal L^{\rm eff}_{b<0} &=& \frac{1}{4 \pi} \frac{p+1}{2 p + 1}AdA,  \ \ \ \ \nu = \frac{p+1}{2p+1} 
\end{eqnarray}
From this, we see that the transformation $b \rightarrow -b$ is equivalent to $\nu \rightarrow 1 - \nu$: particle-hole symmetry is equivalent to flipping the sign of the magnetic field of composite fermions.  But for the zero modes, which are present only when $b < 0$, there is a spectral equivalence between the non-zero energy Landau levels under this transformation.  Hence, we may expect that the corresponding observables are identical in both cases.  

\section{Comparison with Dirac composite fermions}
Next, we compare the results obtained in the previous section with the predictions of Son's Dirac composite fermion theory\cite{Son2015} and show that at the level of mean-field theory considered here, they are identical.  To motivate the Dirac composite fermion theory, let us, for the sake of amusement, consider the problem of a single 2-component Dirac electron in the lowest Landau level.  The corresponding Lagrangian is
\begin{equation}
\mathcal L = i \bar \psi \slashed{D}_A \psi - \frac{1}{8 \pi} AdA + \cdots
\end{equation}
The second term above comes from the parity anomaly.\cite{RedlichDV}  In a simple lattice regulated theory, there is a doubler fermion, which in this case is necessarily massive (since by assumption we have a single, light 2-component Dirac electron), and upon integrating it out, we obtain a level-half Chern-Simons term for $A$.  We motivate the Dirac composite fermion theory by following the usual procedure of flux attachment used in the non-relatvisitic case: we attach two flux quanta by replacing in the Lagrangian $\mathcal L[A] \rightarrow \mathcal L [A+a]$, with $a_{\mu}$ a dynamical $U(1)$ gauge field, and add a level-1/2 Chern-Simons term for $a$:
\begin{eqnarray}
\mathcal L_{cf} &=& i \bar \chi \slashed{D}_{A+a} \chi - \frac{1}{8 \pi} \left(A+a \right)d \left( A+ a \right) + \frac{1}{8 \pi} a da + \cdots  \nonumber \\
&=& i \bar \chi \slashed{D}_{a} \chi - \frac{1}{8 \pi} a d a  + \frac{1}{8 \pi} \left(a-A \right) d \left( a- A \right) + \cdots \nonumber \\
\end{eqnarray}
In the last line above, we again shifted $a_{\mu} \rightarrow a_{\mu} - A_{\mu}$ as before.  This Lagrangian was postulated by Son to describe a particle-hole symmetric lowest Landau level.  The form written above enables us to compare directly the predictions of the HLR theory in the previous section with those of the Dirac theory.  The second term above can be interpreted as  the contribution from a massive Dirac partner whose mass  is much larger than the energy scales of interest.  For the remaining massless fermion, when we tune away from $\nu=1/2$, we fill $(p+1/2)$ Landau levels for either sign of field.  The ``1/2" comes from the zero mode, which is always half-filled.  Thus, 
\begin{eqnarray}
\mathcal L[a,A] &=& \zeta \frac{p+\frac{1}{2}}{4 \pi} ada+ \frac{1}{8 \pi}(a-A)d(a-A) - \frac{1}{8\pi} ada \nonumber \\
&=& \zeta \frac{ \left( p + \frac{1-\zeta}{2} \right)}{4 \pi} ada + \frac{1}{8 \pi}(a-A)d(a-A)
\end{eqnarray}
\

\begin{figure}
\includegraphics[width=3in]{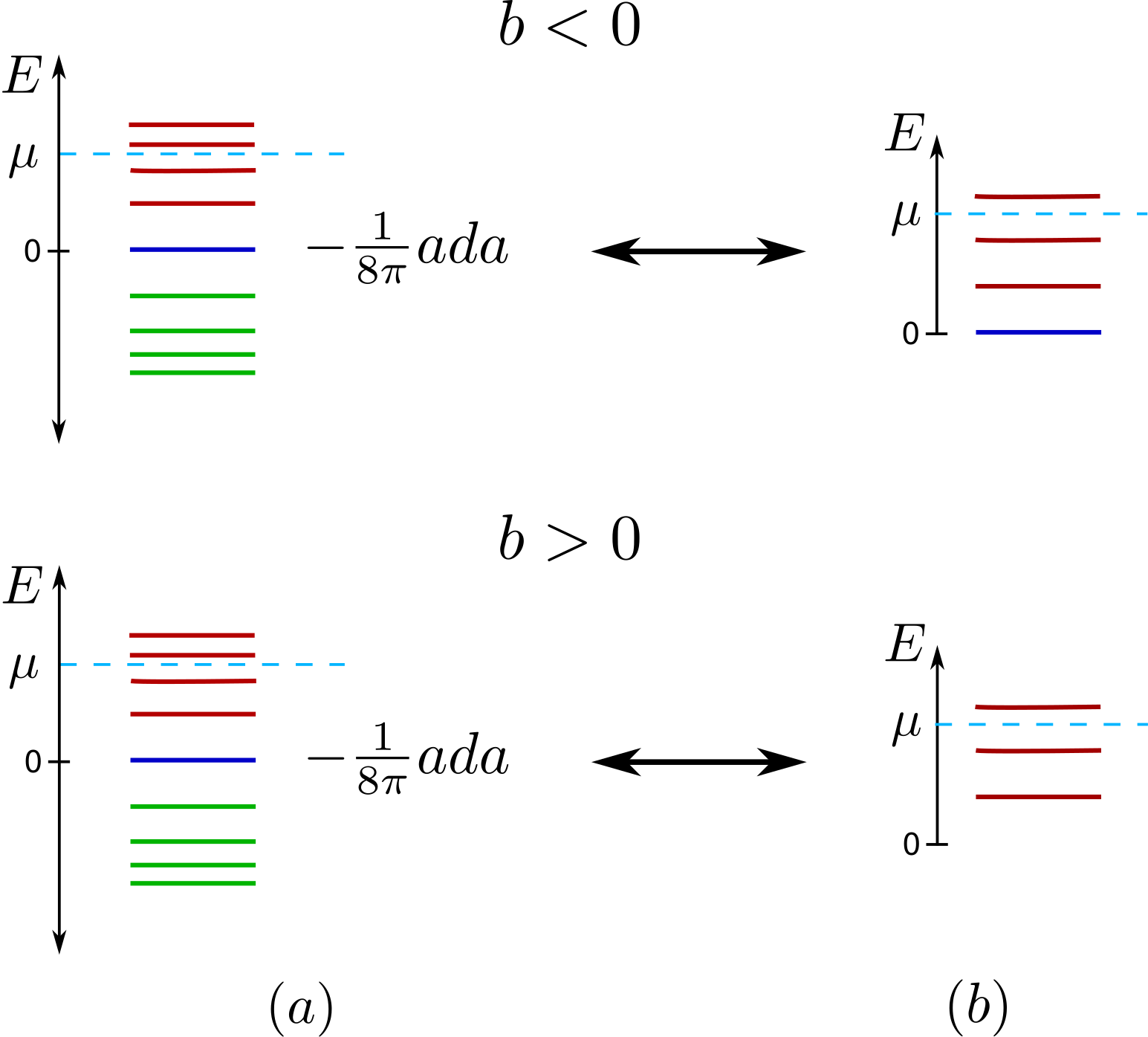}
\caption{(a) A schematic for the response of Dirac composite fermion theory to an applied gate voltage. Energy levels shown with green, red and blue colors represent negative, positive and zero energy Landau levels respectively. The number of filled Landau levels is $p+\frac{1}{2}$ for either sign of the effective magnetic field. Taking into account the contribution of the massive Dirac partner, one gets $p$ filled Landau levels for $b >0$ and $p+1$ Landau levels for $b<0$. This is identical to the response of the HLR theory depicted in (b) as explained in section \ref{resolution_section}.}
\label{Dirac_HLR_eqv_response}
\end{figure}

Consequently, the Dirac theory and the HLR theory described in the previous section have the same response properties.  The massive partner effectively adds another half-filled Landau level that, in conjunction with the Landau levels of the massless Dirac fermion, precisely reproduces the non-relativistic spectrum of Landau levels of HLR theory.  And both satisfy the expectations of particle-hole symmetric electromagnetic response.     

\section{Generalization to even denominator fillings $\nu=\frac{1}{2q}$}
Similar experiments to the one in Ref.~\onlinecite{Pan2019} can be performed in the neighborhood of even denominator fractions at $\nu=1/2q$.  Although such states lack particle-hole symmetry,\cite{PhysRevB.61.R5101} they may exhibit an emergent reflection symmetry in the conductivity tensor.\cite{Shahar1996, ShimshoniSondhiShahar1997} We can generalize our description of $\nu=1/2$ to such states.  At $\nu=1/2q$, the composite fermion Lagrangian is
\begin{equation}
\label{wall}
\mathcal L = \bar f \hat K_{A + a} f + \frac{1}{2q} \frac{1}{4 \pi} ada.
\end{equation}
The above theory can be viewed as arising from 
\begin{equation}
\mathcal L = \bar f \hat K_{A+a+a'} f + \frac{1}{2(q-1)} \frac{1}{4 \pi} a'da' + \frac{1}{2} \frac{1}{4 \pi} a da.
\end{equation}
This can easily be seen by noting that the combination $a-a'$ can be integrated out, since it does not couple to the fermions.  The second form of the Lagrangian is more convenient, as it enables us to view the physics of $\nu=1/2q$ as a half-filled Landau level of composite fermions, which see zero net magnetic field at $\nu = 1/2(q-1)$.  

We proceed as follows.  In mean-field theory, we neglect fluctuations of both $a$ and $a'$ (again this neglect is justified here by our interest in only the gapped phases in the vicinity of $\nu=1/2q$).  Defining 
\begin{equation}
\tilde a = A + a',
\end{equation}
\begin{equation}
\label{berniesanders}
\mathcal L =  \bar f \hat K_{\tilde a + a} f + \frac{\left( \tilde a - A \right) d \left( \tilde a - A \right)}{2(q-1)\dot 4 \pi}  + \frac{1}{2} \frac{1}{4 \pi} a da,
\end{equation}  
and we can now repeat the argument above for $\nu=1/2$ {\it mutatis mutandis}.  Let us start at $\nu=1/2q$, with electron and composite fermion densities set to
\begin{equation}
\langle \bar \psi \psi \rangle = \frac{1}{2q} \frac{B}{2 \pi}  =  \frac{ \tilde b}{4 \pi},
\end{equation}
where $\tilde b = \partial_x \tilde a_y - \partial_y \tilde a_x$.  The second equality above follows directly from analyzing the equations of motion of $a$ and $a'$ in Eq.~\eqref{wall}.  Now, let us vary the chemical potential slightly away from $\nu=1/2q$.  Defining $\mu = \mu_{1/2q} + V$, the composite fermion density changes from $n_{1/2q} = \tilde b/4 \pi$  to
\begin{equation}
\langle \bar f f \rangle = n_{1/2q} + \chi V  = \frac{\tilde b}{4 \pi} + \frac{m}{2 \pi} V = - \frac{b}{4 \pi}.\label{voltage_response}
\end{equation} 
Notice that in deriving this equation, we have assumed that $\tilde b$ is kept fixed. This is so becuase we associate the reflection symmetry at $\nu=1/2q$ with the particle-hole symmetry of the composite fermions constructed by attaching $2(q-1)$ flux quanta to electrons. Experimentally, this requires changing the external magnetic field in proportion to the gate voltage so as to keep $\tilde b$ constant.

From Eq. \eqref{voltage_response}, we infer $V = - (b+\tilde b)/2m$.  Thus, with non-zero $V$, the Lagrangian in Eq.~\eqref{berniesanders} is modified to
\begin{widetext}
\begin{equation}
\label{ivanka}
\mathcal L =  \bar f \left[ \hat K_{a} - \frac{b}{2m} \right] f  + \frac{\left( \tilde a - A \right) d \left( \tilde a - A \right)}{2(q-1) 4 \pi}  + \frac{1}{2} \frac{1}{4 \pi} (a-\tilde a) d(a-\tilde a),
\end{equation}
\end{widetext}
where we shifted $a \mapsto a - \tilde a$.
Note that the first two terms above are precisely of the form discussed in the context of $\nu=1/2$.  Consequently, defining $\zeta = {\rm sgn}(b)$, if $p$ Landau levels are filled for $\zeta = 1$, then $p+1$ Landau levels are filled for $\zeta = -1$.  In such a situation, we may integrate out the gapped fermions in Eq.~\eqref{ivanka} to obtain
\begin{equation}
\mathcal L =  \zeta \frac{p + \frac{1-\zeta }{2}}{4 \pi} a da + \frac{\left( \tilde a - A \right) d \left( \tilde a - A \right)}{2(q-1)\dot 4 \pi}  + \frac{1}{2} \frac{1}{4 \pi} (a-\tilde a) d(a-\tilde a).
\end{equation}
Proceeding as before, we may integrate out $a$ and also $\tilde a$, and obtain 
\begin{eqnarray}
\mathcal L^{eff}_{b>0} &=& \frac{1}{4 \pi} \frac{p}{2q p + 1}AdA,  \nonumber \\
\mathcal L^{eff}_{b<0} &=& \frac{1}{4 \pi} \frac{p+1}{2 q(p+1) - 1}AdA, 
\end{eqnarray}
in direct generalization of the analysis for $\nu=1/2 (q=1)$.  Precisely the same predictions are obtained from recent Dirac composite fermion theories\cite{PhysRevB.98.165137, 2018arXiv180807529W} for $\nu=1/2q$ (see also Ref.~\onlinecite{2017arXiv171204942H}).  In the literature, such a property, which generalizes particle-hole symmetry for $q=1$, is referred to as reflection symmetry.  Exactly as in Dirac based composite fermion theories, the non-relativistic fermions exhibit such reflection symmetry.

\section{Discussion}
In this paper, we have looked at the the response of the half-filled Landau level to a changing chemical potential at a fixed magnetic field. We found that the HLR theory of composite fermions gives rise to a Jain sequence of fractioal quantum Hall states that satisfy particle-hole symmetry. The same conclusion would be obtained from a theory of Dirac composite fermions.  We generalized it to explain the reflection symmetry observed around the compressible states at $\nu=1/2q$ in terms of non-relativistic composite fermions. It is interesting to note that in an alternate situation, where the electron density is kept fixed and the external magnetic field is made variable, HLR and Dirac type theories again produce equivalent response.\cite{PhysRevB.98.115105}

The observations above are, in retrospect, not surprising, if we recall that the properties of free non-relativistic fermions and those of free Dirac fermions in the lowest Landau level are identical.  Both have identical wave functions, and the Dirac spinor of the lowest Landau level has non-zero support only in one of the two components: this why it behaves essentially as a ``spinless" non-relativistic electron.  Imparting flux attachment to both systems should in this case be equivalent.  The conclusions drawn from both ought to be the same, as we have shown.  It is far less clear whether these conclusions remain true in the presence of interactions among the particles.  




\acknowledgments
We thank Y.-B. Kim, S. Kivelson and C. Varma for illuminating discussions.

P.K. and  S.R are supported in part by the 
the DOE Office of Basic Energy Sciences, contract DEAC02-
76SF00515.
M.M. is supported in part by the UCR Academic Senate and the Hellman Foundation.


\bibliography{bigbib}
\bibliographystyle{utphys}

\end{document}